\documentclass[9pt,a4paper]{article}
\usepackage[top=30truemm,bottom=30truemm,left=20truemm,right=20truemm]{geometry}
\usepackage[dvipdfmx]{graphicx}
\usepackage{atbegshi}
\usepackage{amsmath} 
\usepackage{amssymb}
\usepackage{amsfonts}
\usepackage{authblk}
\usepackage{fancyhdr}
\usepackage{comment} 
\usepackage{multicol} 
\usepackage{titlesec} 
\usepackage{ifthen}

\usepackage{color}

\usepackage{secdot}
  \sectiondot{subsection}

\usepackage{lastpage} 
\titleformat*{\section}{\center\normalsize\bfseries} 
\titleformat*{\subsection}{\center\small\bfseries}

\makeatletter 
    
    \@addtoreset{equation}{section}
  \makeatother
  
\begin{document}
\renewcommand{\thefootnote}{\fnsymbol{footnote}} 
\begin{center}
\large{\textbf{$\epsilon$-Expansion in Critical $\phi^3$-Theory on Real Projective Space \\ from Conformal Field Theory}} \\
\end{center}
\begin{center}
\normalsize{Chika Hasegawa$^{1}$ \footnote[1]{\texttt{chika.hasegawa AT rikkyo.ac.jp}}
and Yu Nakayama$^{1}$  \footnote[2]{\texttt{yu.nakayama AT rikkyo.ac.jp}}}  \\
${}^1$\textit{Department of Physics, Rikkyo University, Tokyo 171-8501, Japan} \\
\end{center}
\begin{abstract}
We use a compatibility between the conformal symmetry and the equations of motion to solve the one-point function in the critical $\phi^3$-theory (a.k.a the critical Lee-Yang model) on the $d= 6-\epsilon$ dimensional real projective space to the first non-trivial order in the $\epsilon$-expansion.
It reproduces the conventional perturbation theory and agrees with the numerical conformal bootstrap result.
\end{abstract}

\renewcommand{\thefootnote}{\arabic{footnote}} 
\begin{multicols}{2} 
\section{INTRODUCTION} \label{Introduction}
Unreasonable usefulness of conformal field theories in critical phenomena has been appreciated over many decades first in two dimensions and more recently in higher dimensions. For example, remarkable progress in numerical conformal bootstrap has enables us to predict the critical exponents in the three dimensional critical Ising model in six digits \cite{EPPRSV1}\cite{EPPRSV2}\cite{KPS}\cite{KPSV}. It is, therefore, of our great interest to understand how and why the conformal symmetry, alone or with some additional assumptions, determines the universal nature of critical phenomena.

A hallmark in this direction is the analytic work by Rychkov and Tan \cite{RT}, in which they used the compatibility between the conformal symmetry and the equations of motion in the critical $\phi^4$-theory in $d = 4-\epsilon$ dimensions to reproduce the critical exponents to the first non-trivial order in the $\epsilon$-expansion. More precisely they postulated the following three axioms.
I: The non-trivial fixed point has conformal symmetry.
II: If we take the $\epsilon \rightarrow 0$ limit, correlation functions will  approach the ones in the free theory. 
III: From the equations of motion, a particular primary operator in the free theory behaves as a descendant operator at the non-trivial fixed point.
The use of the equations of motion was further developed in \cite{Nii} to derive critical exponents in the critical $\phi^3$-theory in $d=6-\epsilon$ dimensions by extending the earlier works \cite{Anselmi:1998ms}\cite{Henn:2005mw}\cite{Belitsky:2007jp}. See also \cite{BK}\cite{GGJN}\cite{R}\cite{Y} for more applications of the similar method in various conformal field theories (with defects).

In this paper, we employ the same philosophy to solve the one-point function in the critical $\phi^3$-theory (a.k.a the critical Lee-Yang model) on the $d= 6-\epsilon$ dimensional real projective space to the first non-trivial order in $\epsilon$. 
The application of the technique based on conformal symmetry in the $\phi^3$-theory is non-trivial. The theory is not reflection positive so the constraint coming from the unitarity, which is at the core of the numerical conformal bootstrap, is lacking. Furthermore, on the real projective space, the conformal symmetry is reduced from $SO(d+1,1)$ to $SO(d,1)$. Still the technique of the truncated conformal bootstrap was numerically successful as shown in \cite{N}. 
We will further show that the conformal symmetry is powerful enough to obtain the one-point function analytically, which is additional conformal field theory data on the real projective space. Our analytic results may be compared with the numerical results in good agreement, which shows the validity of the truncated conformal bootstrap approach taken there.

In relation to condensed matter physics, conformal filed theories on unorientable space including a real projective space may be of physical interest. Since a real projective space in even dimensions is not orientable, it seems, at first sight, difficult or even impossible to realize critical systems on such space in our real world and therefore it may appear to be only of academic interest. However, the recent classification of topological phase of matter reveals putting a system on non-orientable manifolds gives us a crucial hint to understand the parity anomaly in the condensed matter physics \cite{Witten:2016cio}. Studying a critical system on real projective space may be one step forward in this direction.

Beyond its theoretical interest in condensed matter physics, there is an unexpected application of conformal field theories on real projective space. In \cite{V}\cite{MNSTW}\cite{NO1}\cite{NO2}\cite{GMT}\cite{LTV}, they realize that the symmetry of bulk local fields in the context of AdS/CFT correspondence may be related to the crosscap Ishibashi states in dual conformal field theories. This is certainly the case in the large $N$ limit, and how to introduce the bulk interactions to preserve the locality is under a lot of discussions. While we do not expect that the critical Lee-Yang model has a weakly coupled bulk dual, but its large $N$ version may be described by a higher spin theories in the bulk. In this sense, the pursuit of the use of the conformal symmetry in the dual field theories on real projective space may be useful.

The rest of the paper is organized as follows:
In section \ref{Conformal Field Theory on Real Projective Space}, based on \cite{N}\cite{NO2}, we summarize the basic facts about conformal field theories on the $d$-dimensional real projective space.
In section \ref{Critical phi3 Theory}, we define the critical $\phi^3$-theory in $d=6-\epsilon$ dimensions from the viewpoint of conformal field theories.
In section \ref{epsilon-Expansions from Conformal Field Theory}, we use a compatibility between the conformal symmetry and the equations of motion to solve the one-point function to the first non-trivial order in $\epsilon$.
In section \ref{Conclusion}, we conclude the paper with some discussions.
In appendix \ref{perturbation result}, as a check of our computation in the main text, we derive the one-point function from the conventional perturbation theory.

\section{CONFORMAL FIELD THEORY ON REAL PROJECTIVE SPACE} \label{Conformal Field Theory on Real Projective Space}
A real projective space $\mathbb{RP}^d$  is defined by an involution $\vec{x} \to -\frac{\vec{x}}{|\vec{x}|^2}$ on the flat Euclidean space $\mathbb{R}^d$ with a $d$-dimensional Cartesian coordinate vector $\vec{x}$. Given a conformal filed theory data on the flat Euclidean space, i.e. the spectrum of the operators and the operator product expansions (OPE) coefficients,  new data on the real projective space is the one-point functions of the primary operators. With the $SO(d,1)$ symmetry, the one-point function of the local operator $O_{i}(x)$ having the scaling dimension $\Delta_{i}$ is given by
\begin{align}
\langle O_{i} (x) \rangle^{\mathbb{RP}^d} = \frac{A_{i}}{(1+x^2)^{\Delta_{i}}}. \label{eq:1pt func for Oi on RPd}
\end{align}

Solving conformal field theories on the real projective space is equivalent to specifying all $A_i$. For example, the two-point functions of scalar primary operators on the real projective space may be expressed as
\begin{align}
& \langle O_{1} (x_{1}) O_{2} (x_{2}) \rangle^{\mathbb{RP}^d} \nonumber \\
& = \frac{(1+x_{1}^{2})^{\frac{-\Delta_{1}+\Delta_{2}}{2}} (1+x_{2}^{2})^{\frac{-\Delta_{2}+\Delta_{1}}{2}} }{|x_{1}-x_{2}|^{2\left( \frac{\Delta_{1}+\Delta_{2}}{2} \right)}} G_{12}(\eta), \label{eq:2pt func for Oi Oj on RPd}
\end{align}
where $G_{12}(\eta)$ has the conformal partial wave decomposition:
\begin{align}
& G_{12}(\eta) = \sum_{i} C_{12}^{\ \ i} A_{i} \eta^{\frac{\Delta_{i}}{2}} \nonumber \\
& \times {}_{2}F_{1} \left( \frac{\Delta_{1} - \Delta_{2} + \Delta_{i}}{2}, \frac{\Delta_{2} - \Delta_{1} + \Delta_{i}}{2}; \Delta_{i}+1-\frac{d}{2}; \eta \right). \label{eq:CPWD 12}
\end{align}
Here $\eta := \frac{(x-y)^2}{(1+x^2)(1+y^2)}$ is called crosscap crossratio.
Note that the sum is taken only over the scalar primary operators appearing in the theory.

There is a consistency condition for the two-point functions on the real projective space. The conformal partial wave decomposition \eqref{eq:CPWD 12} in \eqref{eq:2pt func for Oi Oj on RPd} was obtained in the limit $\vec{x}_{1} \rightarrow \vec{x}_{2}$, but on the real projective space, we may equally well expand in the limit $\vec{x}_{1} \rightarrow - \frac{\vec{x}_{2}}{|\vec{x}_{2}|^{2}}$, which must give the same answer.
As a result, we obtain the constraint equation 
\begin{align}
\left( \frac{1-\eta}{\eta^{2}} \right)^{\frac{\Delta_{1} + \Delta_{2}}{6}} G_{12}(\eta) = \left( \frac{\eta}{(1-\eta)^2}\right)^{\frac{\Delta_{1} + \Delta_{2}}{6}}  G_{12} (1-\eta),
\end{align}
known as the crosscap bootstrap equation.

\section{CRITICAL $\phi^3$-THEORY} \label{Critical phi3 Theory}
In this section, we define the critical $\phi^3$-theory in $d=6-\epsilon$ dimensions from the viewpoint of a conformal field theory. We are going to solve this model in formal power series of $\epsilon^{1/2}$, while we know that such an expansion is only an asymptotic one.

The classical action of critical $\phi^3$-theory in $d=6-\epsilon$ dimensions is 
\begin{align}
S[\phi, g]  = \int \mathrm{d}^{d} x \, \left[ \frac{1}{2} \left( \partial_{\mu} \phi(x) \right)^2 + \frac{g}{3!} \phi^3(x) \right], \label{eq:critical phi3 action}
\end{align}
and the classical equations of motion is 
\begin{align}
\Box_{x} \phi(x) = \frac{g}{2} \phi^2(x), \label{eq:EoM}
\end{align}
here $\Box_{x} := (\partial_{\mu})^2$ is Laplacian in $d$-dimensions. Within the perturbation theory, it is known that $g$ has a non-trivial renormalization group fixed point of order $\epsilon^{1/2}$ and the theory shows conformal invariance at the critical point.
The fixed point value of $g$ is purely imaginary for $\epsilon>0$ and the theory is not reflection positive.

Given this perturbative picture, we postulate the following three axioms:
I: The non-trivial fixed point has conformal symmetry. 
II: If we take the $\epsilon \rightarrow 0$ limit, correlation functions will  approach the ones in the free theory. 
III: From the equations of motion, a particular primary operator in the free theory (i.e. $\phi^2$) behaves as  a descendant operator at the non-trivial fixed point (i.e. $\phi^2$ is a descendant of $\phi$ by acting Laplacian as in \eqref{eq:EoM}).

We have a comment on axiom III. From the purely conformal field theory viewpoint, we, a priori, do not know the magnitude of $g$ at the fixed point nor if this is related to the OPE coefficient $C_{\phi\phi}^{\ \ \phi}$, but from the expectation in the conventional perturbation theory, it is consistent to assume that $g$ is of order $\epsilon^{1/2}$ and so will we in the following. 

As we mentioned in the previous section, our main interest is to determine the one-point functions on the real projective space. In particular, we would like to focus on the one-point function of the lowest dimensional scalar primary operator $\phi$:
\begin{align}
\langle \phi (x) \rangle^{\mathbb{RP}^d} = \frac{A_{\phi}}{(1+x^2)^{\Delta_{\phi}}}. \label{eq:1pt func}
\end{align}
We denote  the scaling dimension of the scalar primary operator $\phi$ as $\Delta_{\phi} = \frac{d-2}{2} + \gamma_{\phi} = 2 - \frac{\epsilon}{2} + \gamma_{\phi},$ where $\gamma_{\phi}$ is the anomalous dimension.

For this purpose, we are going to study its two-point function: 
\begin{align}
\langle \phi (x) \phi (y) \rangle^{\mathbb{RP}^d} = \frac{1}{|x-y|^{2\Delta_{\phi}}} G_{\phi \phi}(\eta), \label{eq:2pt func}
\end{align}
with the conformal partial wave decomposition \cite{NO2}:
\begin{align}
G_{\phi \phi}(\eta) &= \sum_{i} C_{\phi \phi}^{\ \ i} A_{i} \nonumber \\
& \quad \times \eta^{\frac{\Delta_{i}}{2}} {}_{2}F_{1} \left( \frac{\Delta_{i}}{2}, \frac{\Delta_{i}}{2}; \Delta_{i}+1-\frac{d}{2}; \eta \right). \label{eq:CPWD}
\end{align}
For later purposes, we expand (\ref{eq:CPWD}) to the first few terms in $\eta$:
\begin{align}
G_{\phi \phi}(\eta) &= C_{\phi \phi}^{\ \ 1} A_{1} + C_{\phi \phi}^{\ \ \phi} A_{\phi} \frac{1}{\gamma_{\phi}} \nonumber \\
& \quad \times \eta^{\frac{\Delta_{\phi}}{2}} \left[ \gamma_{\phi} + \left( \frac{\Delta_{\phi}}{2} \right)^2 \eta + \mathit{O}(\eta^2) \right] + \cdots, \label{eq:CPWD expand}
\end{align}
by noticing that the scalar OPE is given by $[\phi] \times [\phi] = 1 + [\phi] + [\phi^3] + \cdots$ in the critical $\phi^3$-theory. 

\section{$\epsilon$-EXPANSION FROM CONFORMAL FIELD THEORY} \label{epsilon-Expansions from Conformal Field Theory}
In this section, we apply the axioms of the critical $\phi^3$-theory with conformal invariance to determine the critical exponents and the one-point function on the real projective space to the first non-trivial order in the $\epsilon$-expansion. 

First of all, let us fix the normalization of the correlation functions and use axiom II of the continuity of the correlation functions to the free field theory in the $\epsilon \to 0$ limit. 
We fix the normalization of the two-point function for the lowest dimensional primary operator in the free theory as $\frac{1}{(d-2)S_d}=\frac{1}{4\pi^3}$, where $S_{d}:=\frac{2\pi^{d/2}}{\Gamma(\frac{d}{2})}$ is the surface area of a unit $d$-sphere.
With this normalization, the free field correlation functions on the real projective space are given by
\begin{align}
\langle \phi (x) \rangle^{\mathbb{RP}^d}_{\mathrm{free}} &= 0, \label{eq:1pt func free} \\
\langle \phi^2 (x) \rangle^{\mathbb{RP}^d}_{\mathrm{free}} &= \frac{1}{4 \pi^3} \frac{1}{(1+x^2)^4}, \label{eq:phi2 correlation func free} \\
\langle \phi (x) \phi (y) \rangle^{\mathbb{RP}^d}_{\mathrm{free}} &= \frac{1}{4 \pi^3} \frac{1}{|x-y|^4} \left[ 1 + \left( \frac{\eta}{1-\eta} \right)^2 \right], \label{eq:2pt func free} \\
\langle \phi^2 (x) \phi^2 (y) \rangle^{\mathbb{RP}^d}_{\mathrm{free}} &= \left( \frac{1}{4 \pi^3} \right)^2 \frac{1}{|x-y|^8} \nonumber \\ 
& \quad \times \left[ 2 \cdot \left[ 1 + \left( \frac{\eta}{1-\eta} \right)^2 \right]^2 + \eta^4 \right], 
 \label{eq:phi2 phi2 correlation func free}
\end{align}
where we mean by $\langle \cdots \rangle^{\mathbb{RP}^d}_{\mathrm{free}}$ that the expectation values are evaluated in the free theory with $\epsilon=0$.
The above correlation functions are obtained by using the method of image under the involution $\vec{x} \to -\frac{\vec{x}}{|\vec{x}|^2}$. 

We now demand that \eqref{eq:2pt func} approaches \eqref{eq:2pt func free} in the $\epsilon \to 0$ limit. For this to be possible, as more explicitly seen in \eqref{eq:CPWD expand}, we need to require:
\begin{align}
C_{\phi \phi}^{\ \ 1} A_{1} &= \frac{1}{4 \pi^3} +  \mathit{O}(\epsilon), \\
C_{\phi \phi}^{\ \ \phi} A_{\phi} \frac{1}{\gamma_{\phi}} &= \frac{1}{4 \pi^3} + \mathit{O}(\epsilon). \label{eq:relation}
\end{align}
This is our first main result.

The next goal is to determine each piece of the left hand side of \eqref{eq:relation} separately. For this purpose, we combine axiom II and III in the correlation functions and take the $\epsilon \to 0$ limit. 
Let us begin with the one-point function.
Axiom III means that we can use the classical equations of motion
\begin{align}
\langle \Box_{x} \phi (x) \rangle^{\mathbb{RP}^d} = \frac{g}{2} \langle \phi^2 (x) \rangle^{\mathbb{RP}^d}, \label{eq:Laplacian acting 1pt func}
\end{align}
inside the one-point function to derive the consistency condition to the first non-trivial order in the $\epsilon$-expansion.
By acting Laplacian on  \eqref{eq:1pt func} and comparing it with \eqref{eq:phi2 correlation func free} and \eqref{eq:Laplacian acting 1pt func}  to the first non-trivial order in the $\epsilon$-expansion
\begin{align}
(\mathrm{LHS}\eqref{eq:Laplacian acting 1pt func}) &= \Box_{x} \langle \phi (x) \rangle^{\mathbb{RP}^d} \sim - \frac{24A_{\phi}}{(1+x^2)^4}, \\
(\mathrm{RHS}\eqref{eq:Laplacian acting 1pt func}) &\sim \frac{g}{2} \langle \phi^2 (x) \rangle^{\mathbb{RP}^d}_{\mathrm{free}} = \frac{g}{2} \frac{1}{4 \pi^3} \frac{1}{(1+x^2)^4},
\end{align}
we obtain 
\begin{align}
A_{\phi} = - \frac{g}{48} \frac{1}{4 \pi^3} + \mathit{O}(\epsilon). \label{eq:1pt func coefficient}
\end{align}
This result may be also derived from the conventional perturbation theory by evaluating a Feynman diagram on the real projective space (see Appendix \ref{perturbation result}).

To go further, we study the two-point functions with axiom II and III. We apply the classical equations of motion twice in the two-point function:
\begin{align}
\langle \Box_{x} \phi (x) \Box_{y} \phi (y) \rangle^{\mathbb{RP}^d} = \frac{g^2}{4} \langle \phi^2 (x) \phi^2 (y) \rangle^{\mathbb{RP}^d}. \label{eq:twice Laplacian acting 2pt func}
\end{align}
We now evaluate the left hand side and the right hand side separately to the first non-trivial order in the $\epsilon$-expansion to derive additional necessary conditions in order for the critical exponents to be compatible with the conformal symmetry. 
We focus on the limit when $x$ approaces $y$ (i.e. $\eta \to 0$) by using the operator product expansion. Expanding with respect to $\eta$, the left hand side becomes
\begin{align}
&(\mathrm{LHS}\eqref{eq:twice Laplacian acting 2pt func}) = \Box_{x} \Box_{y} \langle \phi (x)  \phi (y) \rangle^{\mathbb{RP}^d} \nonumber \\
&\quad \sim \frac{1}{4\pi^3} \Box_{x} \Box_{y} |x-y|^{-2\Delta_{\phi}} \nonumber \\
&\qquad \times \left[ 1 + \eta^{\frac{\Delta_{\phi}}{2}} \left( \gamma_{\phi} + \left( 1 - \frac{\epsilon}{2} + \gamma_{\phi} \right) \eta + \mathit{O}(\eta^2) \right)   + \cdots \right], \label{eq:LHS twice Laplacian acting 2pt func}
\end{align}
while the right hand side becomes
\begin{align}
&(\mathrm{RHS}\eqref{eq:twice Laplacian acting 2pt func}) \sim \frac{g^2}{4}  \langle \phi^2 (x) \phi^2 (y) \rangle^{\mathbb{RP}^d}_{\mathrm{free}} \nonumber \\
&\quad = \frac{g^2}{4} \left( \frac{1}{4 \pi^3} \right)^2 \frac{1}{|x-y|^8} \left[ 2 + 4 \eta^2 + \mathit{O}(\eta^3) \right], \label{eq:RHS twice Laplacian acting 2pt func}
\end{align}
in the first non-trivial order in the $\epsilon$-expansion.
The equality must be satisfied as a power series expansion with respect to $\eta$. We will pay attention to the terms of order $\eta^0$, $\eta$, and $\eta^2$ because the $\mathit{O}(\eta^3)$ term has a contribution from higher dimensional primary operators on the left hand side, which we are not interested in.

At order $\eta^0$, directly acting the Laplacian twice on the left hand side of \eqref{eq:twice Laplacian acting 2pt func} (see also \eqref{eq:LHS twice Laplacian acting 2pt func}), we obtain
\begin{align}
& \Box_{x} \Box_{y} |x-y|^{-2\Delta_{\phi}} \nonumber \\
& = 2\Delta_{\phi} ( 2\Delta_{\phi} + 2 - d ) ( 2\Delta_{\phi} + 2 ) ( 2\Delta_{\phi} + 4 - d ) \nonumber \\
& \quad \times |x-y|^{-2\Delta_{\phi}-4} \nonumber \\
& \sim 2 \cdot 3 \cdot 4^2 \gamma_{\phi} |x-y|^{-8},
\end{align}
where we take $\epsilon \to 0$ in the last line. 
On the other hand, the coefficient of this term must agree with the right hand side (multiplied by $4\pi^3$) to the first non-trivial order in $\epsilon$, i.e. $\frac{g^2}{2 \cdot 4 \pi^3}$ at order $\eta^0$. Thus, we obtain 
\begin{align}
\gamma_{\phi} = \frac{g^2}{3 \cdot 4^3 \cdot 4 \pi^3}. \label{eq:gamma form 1st term}
\end{align}
The computation here is essentially same as in the case of flat Euclidean space obtained in \cite{Unpublished}\cite{Nii}.

The comparison at order $\eta$ and $\eta^2$ is more involved. We need to compute the following term with $n=0$ and $1$:
\begin{align}
& \Box_{x} \Box_{y} \left( |x-y|^{-2\Delta_{\phi}}  \eta^{\frac{\Delta_{\phi}}{2} + n} \right) \nonumber \\
& = \left[ a_{(n)} + b_{(n)} \eta + \mathit{O}(\eta^2) \right] |x-y|^{-2\Delta_{\phi}-4}  \eta^{\frac{\Delta_{\phi}}{2} + n},
\end{align}
\begin{align}
a_{(n)} &:= ( \Delta_{\phi} - 2n )( \Delta_{\phi} + 2 - d - 2n ) \nonumber \\
& \quad \times ( \Delta_{\phi} + 2 - 2n )( \Delta_{\phi} + 4 - d - 2n ), \label{eq:coefficient a} \\
b_{(n)} &:= ( \Delta_{\phi} + 2n )( \Delta_{\phi} + 2 - d - 2n ) \nonumber \\
& \quad \times ( \Delta_{\phi} + 2 - 2n )( \Delta_{\phi} - 4n ) \nonumber \\
& \quad - 2d ( \Delta_{\phi} + 2n )( \Delta_{\phi} - 2n )( \Delta_{\phi} + 2 - 2n ) \nonumber \\ 
& \quad + \mathit{O}(x^2). \label{eq:coefficient b}
\end{align}
that appears in the expansion \eqref{eq:LHS twice Laplacian acting 2pt func}.

With this result, let us first compare the term of order $\eta$. Actually, this term vanishes on the right hand side of \eqref{eq:twice Laplacian acting 2pt func} to the first non-trivial order in the $\epsilon$-expansion (see \eqref{eq:RHS twice Laplacian acting 2pt func}), and so must be on the left hand side.  Indeed, this is the case because the $a_{(0)}$ term is the only contribution at order $\eta$, but it is of order $\epsilon$:
\begin{align}
a_{(0)} &= \Delta_{\phi} (\Delta_{\phi}+2-d) (\Delta_{\phi}+2) (\Delta_{\phi}+4-d) \nonumber \\
&= \Delta_{\phi} (\Delta_{\phi}+2-d) (\Delta_{\phi}+2)\left( \frac{\epsilon}{2} + \gamma_{\phi} \right) = \mathit{O}(\epsilon), \nonumber
\end{align}
and it is further multiplied by $\gamma_\phi = \mathit{O}(\epsilon)$ from the expansion coefficient that appeared in \eqref{eq:LHS twice Laplacian acting 2pt func}.
Thus, the term of order $\eta$ does not appear as expected.

Now we compare the term of order $\eta^2$ to the first non-trivial order in the $\epsilon$-expansion. There are two contributions to this order in the left hand side. 
(i) From the $b_{(0)}$ term: since there is  an $\mathit{O}(\epsilon)$ prefactor in $\gamma_{\phi}$, we have to focus on the $\mathit{O}(1)$ term that appeared in $b_{(0)}$ to compare with the right hand side which is of order $g^2 = \mathit{O}(\epsilon)$. (ii) From the $a_{(1)}$ term: since there is a $(1+\mathit{O}(\epsilon))$ prefactor of $(1-\frac{\epsilon}{2}+\gamma_{\phi})$, we have to focus on the $\mathit{O}(1)$ and $\mathit{O}(\epsilon)$ terms that appeared in $a_{(1)}$.
Thus, we approximate
\begin{align}
b_{(0)} &= 2 (\Delta_{\phi})^2 (\Delta_{\phi}+2-d) (\Delta_{\phi}+2) \nonumber \\
& \quad -2d(\Delta_{\phi})^2(\Delta_{\phi}+2-d) + \mathit{O}(x^2) \nonumber \\
&\sim 2 \cdot 4^2 = \mathit{O}(1), \\
a_{(1)} &= (\Delta_{\phi}-2)(\Delta_{\phi}-d)(\Delta_{\phi})(\Delta_{\phi}+2-d) \nonumber \\
&= \left( -\frac{\epsilon}{2} + \gamma_{\phi} \right)(\Delta_{\phi}-d)(\Delta_{\phi})(\Delta_{\phi}+2-d) \nonumber \\
&\sim 4^2 \left( -\frac{\epsilon}{2} + \gamma_{\phi} \right) = \mathit{O}(\epsilon).
\end{align}
Combining the above two contributions, the coefficient of order $\eta^2$ to first non-trivial order in the left hand side is obtained as $b_{(0)}\gamma_{\phi} + a_{(1)} \sim 2 \cdot 4^2\gamma_{\phi} + 4^2(-\frac{\epsilon}{2}+\gamma_{\phi}) = 4^2(-\frac{\epsilon}{2}+3\gamma_{\phi})$, and  this should agree with the right hand side (multiplied by $4\pi^3$), i.e. $\frac{g^2}{4 \pi^3}$ at order $\eta^2$.

In this way, the matching of order $\eta^2$ gives rise to a non-trivial condition on the critical exponents 
\begin{align}
\gamma_{\phi}=\frac{\epsilon}{6}+\frac{g^2}{3 \cdot 4^3 \pi^3}, \label{eq:gamma form 2nd term}
\end{align}
which agrees with the constraint coming from the study of the three-point functions in the flat Euclidean space (see (4.16) in \cite{Nii}).

With all these constraints from axiom II and III, we can completely specify the conformal field theory data on the real projective space. From \eqref{eq:gamma form 1st term} and \eqref{eq:gamma form 2nd term}, we obtain
\begin{align}
g^{2} &= - \frac{2 \cdot 4^3 \pi^3}{3} \epsilon +  \mathit{O}(\epsilon^2),  \label{eq:coupling} \\
\gamma_\phi& = -\frac{1}{18} \epsilon +  \mathit{O}(\epsilon^2), \label{eq:anomalous dimension}
\end{align}
and  determine
\begin{align}
C_{\phi \phi}^{\ \ \phi} A_{\phi}= -\frac{1}{72 \pi^3} \epsilon +  \mathit{O}(\epsilon^2). \label{eq:relation CA}
\end{align}

The last quantity $C_{\phi \phi}^{\ \ \phi} A_{\phi}$ may be computed in the numerical truncated conformal bootstrap on the real projective space as studied in \cite{N} with the help of the bulk data as also numerically studied \cite{Gliozzi:2013ysa}\cite{Gliozzi:2014jsa} by using the numerical truncated conformal bootstrap. We have checked that at $\epsilon =0.05$, the truncated bootstrap only with the first two-terms $[\phi] \times [\phi] = 1 + [\phi] + [\epsilon']$ gives $(4\pi^3)C_{\phi \phi}^{\ \ \phi} A_{\phi} \sim -0.003$ in good agreement (within $10\%$ error) with the $\epsilon$-expansion obtained in the above computation.

\section{CONCLUSION} \label{Conclusion}
We have studied the one-point function of the lowest dimensional scalar primary operator in critical $\phi^3$-theory (a.k.a the critical Lee-Yang model) on the $d=6-\epsilon$ dimensional real projective space by using a compatibility between the conformal symmetry and the classical equations of motion to the first non-trivial order in $\epsilon$.
Our results are in complete agreement with both conventional perturbation theory and numerical conformal bootstrap.
We therefore conclude that the critical $\phi^3$-theory can be defined uniquely through the conformal symmetry and the equations of motion, not only on the flat Euclidean space but also on the non-trivial space to the first order in the $\epsilon$-expansion. It is an interesting open question if such a characterization may go beyond the first order in the $\epsilon$-expansion or even non-perturbatively. 

As a future direction, we may extend our computation in the other critical phenomena such as the critical Ising model or $O(N)$ models in various dimensions. In particular, our method should be directly applicable to the recently discussed $O(N)$ symmetric fixed point in $d=6-\epsilon$ dimensions \cite{LGK}. 

In addition, it is important to determine the conformal field theory data on the real projective space completely. In order to do that, we need to compute one-point functions of all the scalar primary operators beyond the only lowest one which has been studied in this paper. Since it becomes harder and harder to determine the one-point functions of higher dimensional operators from the numerical conformal bootstrap \cite{N}, it is important to develop alternative method such as the one pursued in this paper.  At the same time, the $\epsilon$-expansion must break down for sufficiently higher dimensional operators because of the non-perturbative operator mixing, and it is interesting to see how this breakdown becomes manifest in this  approach based on conformal field theory.

\section*{ACKNOWLEDGMENTS}
Y.~N. is supported in part by Rikkyo University Special Fund for Research.

\section*{APPENDIX}
\appendix
\section{CONVENTIONAL PERTURBATION THEORY} \label{perturbation result}
In this appendix, we derive the one-point function on the real projective space from the conventional perturbation theory.
The classical action of critical $\phi^3$-theory in $d=6-\epsilon$ dimensions is 
\begin{align}
S[\phi, g] = \int \mathrm{d}^{d} x \, \left[ \frac{1}{2} \left( \partial_{\mu} \phi(x) \right)^2 + \frac{g}{3!} \phi^3(x) \right],
\end{align}
and the model is defined by the path integral
\begin{align}
 Z[g]  = \int \mathcal{D} \phi \, e^{-S[\phi,g]}, \ 
\end{align}
with the perturbative expansions in $g$. At $g=0$, the free field correlation functions on the real projective space are given by
\begin{align}
\langle \phi (x) \rangle^{\mathbb{RP}^d}_{\mathrm{free}} &= 0, \label{eq:1pt func free appendix} \\
\langle \phi^2 (x) \rangle^{\mathbb{RP}^d}_{\mathrm{free}} &= \frac{1}{4 \pi^3} \frac{1}{(1+x^2)^4}, \label{eq:phi2 correlation func free appendix} \\
\langle \phi (x) \phi (y) \rangle^{\mathbb{RP}^d}_{\mathrm{free}} &= \frac{1}{4 \pi^3} \frac{1}{|x-y|^4} \left[ 1 + \left( \frac{\eta}{1-\eta} \right)^2 \right], \label{eq:2pt func free appendix}
\end{align}
where $\eta := \frac{(x-y)^2}{(1+x^2)(1+y^2)}$ is crosscap crossratio.
Using the perturbative expansions, we obtain
\begin{align}
\langle \phi (x) \rangle^{\mathbb{RP}^d} &= \langle \phi (x) \rangle^{\mathbb{RP}^d}_{\mathrm{free}} \nonumber \\ &\quad - \frac{g}{3!} \int \mathrm{d}^{d} y \, \langle \phi (x) \phi^3 (y) \rangle^{\mathbb{RP}^d}_{\mathrm{free}} + \mathit{O}(g^2), \\
&= - \frac{g}{2} \int \mathrm{d}^{d} y \, \langle \phi (x) \phi (y) \rangle^{\mathbb{RP}^d}_{\mathrm{free}} \langle \phi^2 (y) \rangle^{\mathbb{RP}^d}_{\mathrm{free}} + \mathit{O}(g^2), \label{eq:1pt func perturbation appendix}
\end{align}
where the integral range is $0\leq|y|\leq1$ and $d=6$ as $\epsilon \to 0$.
In the second line, we use the standard Wick contraction.
After setting $\vec{x}$ to $\vec{0}$ and plugging in \eqref{eq:phi2 correlation func free appendix} and \eqref{eq:2pt func free appendix} for \eqref{eq:1pt func perturbation appendix}, we obtain
\begin{align}
\langle \phi (0) \rangle^{\mathbb{RP}^d} = - \frac{g}{48} \frac{1}{4 \pi^3} + \mathit{O}(g^2). \label{eq:1pt func coefficient perturbation}
\end{align}
Since $\langle \phi (0) \rangle^{\mathbb{RP}^d}$ equals to $A_{\phi}$ in conformal filed theory on the real projective space (see \eqref{eq:1pt func}), this perturbative result agrees with \eqref{eq:1pt func coefficient} which is obtained by using the axioms in the critical $\phi^3$-theory with conformal symmetry discussed in the main text.

\end{multicols}
\hrulefill
\begin{multicols}{2}

\end{multicols}

\end{document}